# Ride Acceptance Behaviour Investigation of Ride-sourcing Drivers Through Agent-based Simulation

Farnoud Ghasemi*[1], Peyman Ashkrof [2], Rafal Kucharski[3]

[1] PhD Candidate, Mathematics and Computer Science, Jagiellonian University, Poland

[2] PhD Candidate, Delft University of Technology, Netherlands

[3] Assistant Professor, Jagiellonian University, Poland

## SHORT SUMMARY

Ride-sourcing platforms such as Uber and Lyft offer drivers (i.e., platform suppliers) considerable freedom of choice in multiple aspects. At the operational level, drivers can freely accept or decline trip requests that can significantly impact system performance in terms of travellers' waiting time, drivers' idle time and income. Despite the extensive research into the supply-side operations, the behavioural aspects, particularly drivers' ride acceptance behaviour remains so far largely unknown. To this end, we reproduce the dynamics of a two-sided mobility platform on the road network of Delft using an agent-based simulator. Then, we implement a ride acceptance decision model enabling drivers to apply their acceptance strategies. Our findings reveal that drivers who follow the decision model, on average, earn higher income compared to drivers who randomly accept trip requests. The overall income equality between drivers with the acceptance decision is higher and travellers experience lower waiting time in this setting.

## INTRODUCTION

Ride-sourcing companies – also known as Transportation Network Companies (TNC) - such as Uber, Lyft, and Didi are two-sided mobility platforms working based on gig economy business models. In this context, flexible working conditions are the primary motivation for drivers who are considered as the platform suppliers to join the ride-sourcing platforms (Hall and Krueger, 2018). Importantly, this flexibility comes with freedom of choice which significantly affect drivers' income and system performance. In accordance with the literature, the main decisions of the drivers at the operational and tactical levels are ride acceptance choice, relocation strategies, and selecting working shift and area (Ashkrof et al., 2020). While the studies on the supply side of ride-sourcing systems cover a wide range of topics, most of them assume that the service is operated by fully compliant drivers or fully automated vehicles. Nevertheless, drivers interacting with the system have very distinct behaviour which stems from their preferences and strategies, and do not necessarily follow the platform's objectives. Besides, automated vehicles are still not present in the current transport systems.

There are few studies examining the ride acceptance decision and its impact on the system performance. Xu et al. (2018) reveal that approximately 40% of ride-sourcing requests are dismissed and receive no response from any driver, which gives rise to further implications for the system performance. Indeed, the low acceptance rate increases the number of unfulfilled requests and the travellers' waiting time, and eventually negatively affects the level of service, platform profit, and drivers' income. Despite several measures taken by platforms like economic incentives to tackle the negative effect of the low acceptance rate, the problem remains unsolved.

This calls for further studies on the nature of the ride acceptance decision and its consequences on the system. To this end, in this study, we first reproduce the dynamics of a two-sided mobility platform on the road network of Delft in the Netherlands using our MaaSSim agent-based simulator (Kucharski and Cats, 2020). We define two main classes of agents: drivers and travellers. Next, we implement an acceptance decision model based on a binary logit model derived from a recent stated preference experiment. We investigate the effects of the ride acceptance decision on the system performance in the simulation framework and then evaluate several key performance indicators such as travellers' waiting time as well as drivers' idle time and income.

## METHODOLOGY

Considering the characteristics of ride-sourcing parties, we utilize our agent-based simulator, MaaSSim, to efficiently reproduce the dynamics of the ride-sourcing environment. Agent-based Modelling (ABM) is the most used computational and modelling framework to simulate independent decision-makers (agents) with different tastes and preferences, as well as to capture the potential interactions between them (Macal and North, 2011). We define two main classes of agents. On the demand side, we have travellers requesting rides from an origin to a destination at a certain time, and on the supply side, there are drivers offering rides to them. Since supply and demand are microscopic, each agent explicitly represents a single driver or traveller, as well as its movements in time and space (road network graph). Besides, agents can have various behavioural characteristics through user-defined decision modules. Apart from this, the platform is defined as an intermediate agent linking the demand to supply through a matching algorithm. Notably, we use the "*first-dispatch*" protocol for matching, which simply pairs the traveller with the closest idle driver who is predicted to have the shortest travel time (Yan and Korolko, 2019).

For this study, we developed a ride acceptance decision module for the MaaSSim simulator. This module is developed based on a recent stated preference experiment conducted by Ashkrof et al., (2021), which investigates the effect of various attributes on the ride acceptance choice of drivers. Accordingly, the ride acceptance choice set $C=\{a,r\}$ is comprised of alternative *a*, *request-acceptance*, and alternative *r*, *request-rejection*, with a set of influential attributes $I$. We calculate the corresponding utility and probability of each alternative based on the Logit model formulation as follows:

$$V_{na} = \sum_{i \in I} \beta_{nai} \cdot x_{nai} \quad (1) \qquad P_{na} = \frac{Exp(V_{na})}{Exp(V_{na}) + 1} \quad (2)$$

$$V_{nr} = 0 \quad (3) \qquad P_{nr} = \frac{1}{Exp(V_{na}) + 1} \quad (4)$$

Where eq. (1) presents the systematic utility of alternative *a* perceived by the driver *n* as a function of parameters ($\beta_i$) and attributes ($x_i$), eq. (2) represents the probability of the alternative *a* to be chosen. Due to the binary characteristic of the ride acceptance choice set, the model parameters were only estimated for the *request-acceptance* alternative. Therefore, the systematic utility of alternative *r* ($V_{nr}$) is equal to zero as shown in eq. (3) with the corresponding probability indicated in eq. (4).

The attributes and the parameters considered for the ride acceptance decision model coupled with their definitions are illustrated in table 1. The beta values were estimated for the *request-acceptance* alternative, which means any attribute with a negative beta value is decreasing the chance of acceptance. While the model includes five main attributes, Alternative Specific Constant (ASC) captures the effect of unobserved attributes on the utility of alternatives.

Table 1: Decision model parameters

| Attributes | $\beta$ value | Definition |
|---|---|---|
| ASC | 1.5 | Alternative specific constant with the value equal to 1 |
| Pick-up time [min] | -0.0491 | The travel time from the driver's location to the traveller's spot |
| Waiting time [min] | -0.0173 | The duration between the last drop-off and the incoming ride |
| Time1_Loc | -0.265 | Dummy variable which is equal to 1 if and only if the driver is at the beginning (first one-third) of his/her working shift and city centre. Otherwise, it is equal to 0. |
| Req_Long_Rate_Declined | 0.0909 | Equal to the product of request type (1 if it is non-shared and 0 if it is shared), trips distance (1 if it is greater than 80% of all trip distances, otherwise 0), traveller's rating out of 5 and the previous request status of the driver (1 if it was declined, otherwise 0) |

## RESULTS

We start our experiments by setting the road network and arbitrary parameters. For illustrative purposes, we simulate a medium-size city, Delft in the Netherlands. The simulation time for each experiment is 5 hours and all simulated rides are non-shared type (e.g., UberX). We assume a hypothetical central area for the city with a geographical centre and 1.5 [km] radius. While the vehicle speed is fixed to 18 [km/h] in the central area, the vehicle speed on the rest of the road network is constant and 36 [km/h]. This speed difference helps us reproduce the congestion in the city centre. The number of driver and traveller agents are set to 20 and 500, respectively – which we found to be a condition where drivers do not remain idle for a long time and travellers do not wait a lot to be picked up. Drivers are distributed over the road network synthetically with random starting positions at the graph node level. The trip fare calculated by the platform (e.g., Uber) is based on a base fare and an additional fare per kilometre depending on the city. However, for simplicity, we assume a trip fare of 2 [€/km] which is the calculated average fare based upon the Uber price estimator (Uber Technologies Inc., 2021).

Subsequently, we define two classes of drivers, (i) behavioural class, drivers who employ the ride acceptance model, (ii) random class, drivers who accept ride requests randomly with a predefined probability. Behavioural drivers rely on our decision model to decide about incoming ride requests, while random drivers randomly accept the requests. For the ease of comparison between the two models, we assume the average acceptance probability of both classes is the same. Thus, we have two groups of drivers with an equal acceptance probability, behavioural and random, who employ and do not employ the decision model, respectively. We illustrate the decision model and the role of attributes in Fig. 1, which shows that the pick-up time is the most influential attribute on the ride acceptance behaviour.

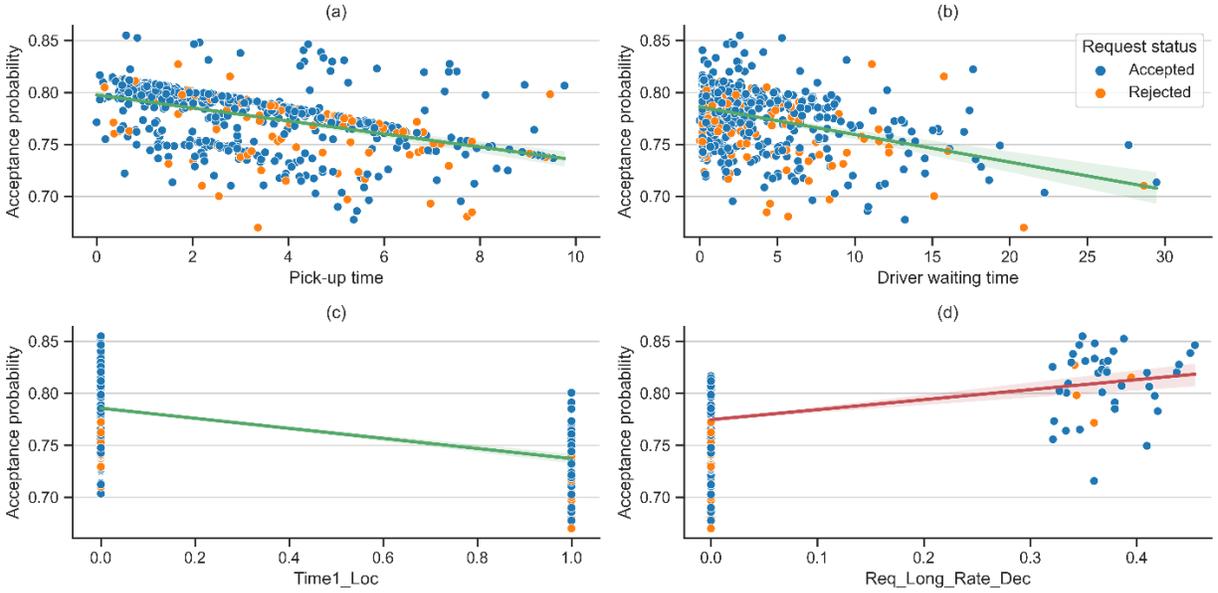

**Figure. 1.** Sensitivity of the ride acceptance probability to the attributes of the ride acceptance model (defined in table 1). While *pick-up time* **(a)**, *waiting time* **(b)** and *Time1_Loc* attributes **(c)** have a negative impact on the acceptance probability, *Req_Long_Rate_Declined* attribute **(d)** positively affects the acceptance probability. Furthermore, *pick-up time* has the most significant impact, however, the model remains highly non-deterministic (visible as the blue and orange dots scattered around the plots).

To inspect the impact of the ride acceptance model on the system, we run the simulations with 20 random drivers and 500 travellers. After 10 replications of the same setup (100% random), we gradually replace random drivers with behavioural ones. Eventually, we reach the setup with 100% behavioural (in 11 steps, each replicated 10 times). Fig 2. presents the strong trend in the driver income (a) and traveller waiting time (b).

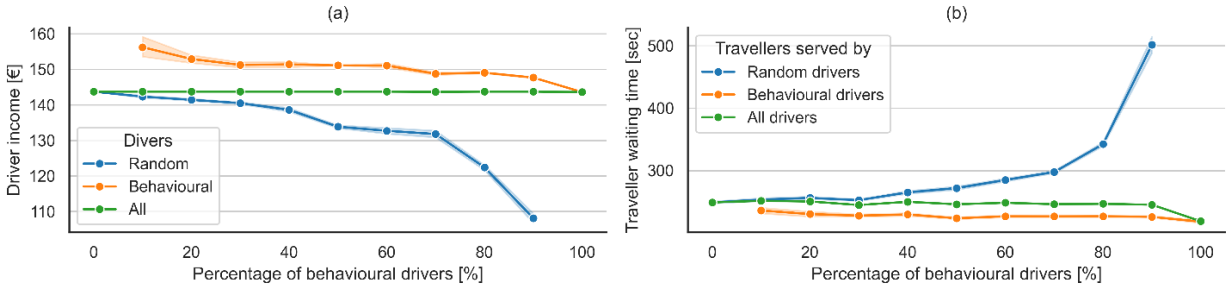

**Figure. 2.** Driver income **(a)** and traveller waiting time **(b)** varying with the share of behavioural drivers in service. **(a)** In general, behavioural drivers' (orange) income is greater than random drivers' (blue) income. Notably, the income of random drivers decreases as the share of behavioural drivers increases. Furthermore, the average income of all drivers (green) remains constant. **(b)** On the whole, travellers served by random drivers (blue) are waiting more than those who travel with behavioural drivers (orange). As the share of behavioural drivers increases, the waiting time of travellers served by random drivers also increases. Moreover, the average waiting time of all travellers (green) decreases with the increasing share of behavioural drivers.

Behavioural drivers and travellers served by them appear to be advantaged. Moreover, they suffer from inequalities less than random drivers (Fig. 3). This emphasizes the importance of decision models in simulation results, particularly inequality issues in ride-sourcing (Bokányi and Hannak, 2020).

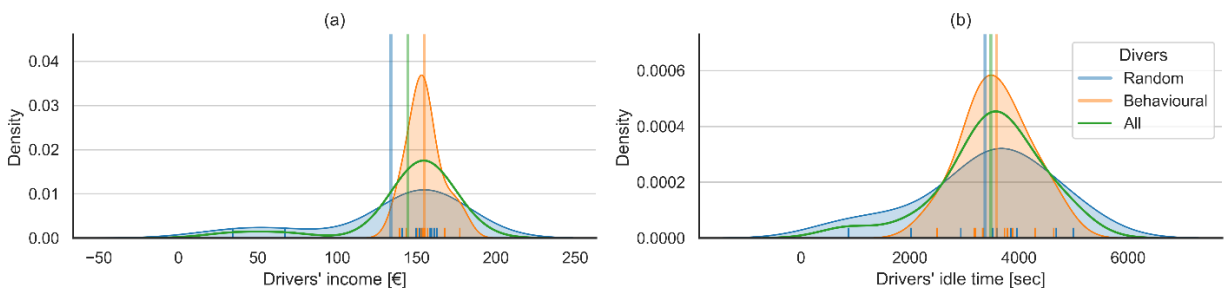

**Figure. 3.** Distribution of driver income **(a)** and driver waiting time **(b)** with 50% behavioural and 50% random drivers. There is a considerable income difference between the two groups of drivers. Not only the behavioural drivers are earning more, but also with less variability compared to random drivers. Besides, greater inequality can be observed in random drivers' idle time. However, the averages for both groups are almost the same.

## CONCLUSIONS

In this research, we investigated the ride acceptance behavior of the ride-sourcing drivers in our agent-based simulation framework. We reproduced the ride-sourcing environment in MaaSSim simulator for the road network of Delft city in the Netherlands, where we implemented a novel binary logit model for the ride acceptance decision. Our results show that pick-up time has the most substantial impact on drivers' decisions among the attributes. However, the model is less sensitive to the other attributes like drivers' waiting time and location, trip distance, travellers' rating, etc.

Furthermore, we found that introduction of the ride acceptance behavior for the system can significantly affect the platform performance. Two classes of drivers (behavioral drivers who follow the ride acceptance behaviour model and random drivers who randomly accept/reject the incoming requests) and travellers served by them are dealing with very different experiences in the same platform. In addition, this difference is growing as the share of behavioral drivers in the system increase. While, on the average, behavioral drivers earn more, travellers served by random drivers wait longer. Moreover, the implementation of the decision model equalizes the incomes among the drivers. Thus, we advocate for including this behavior in future ride-sourcing simulations.

## ACKNOWLEDGMENT

This research was funded by National Science Centre in Poland program OPUS 19 (Grant Number 2020/37/B/HS4/01847) and by Jagiellonian University under the program Excellence Initiative: Research University (IDUB).

This work was supported by the H2020 European Research Council under Grant 804469 and by the Amsterdam Institute for Advanced Metropolitan Solutions.